\documentclass[12pt]{iopart}

\usepackage{graphicx}% Include figure files
\usepackage{bm}% bold math
\newcommand{\veps}{\varepsilon}

\begin{document}

\title{An order-$N$ electronic structure theory with generalized eigen-value equations
and its application to a ten-million-atom system
}

\author{T. Hoshi$^{1,2}$, S. Yamamoto$^{3}$, T. Fujiwara$^{4}$, T. Sogabe$^{5,2}$, S.-L. Zhang$^{6,2}$}

\address{(1) Department of Applied Mathematics and Physics, 
Tottori University, Tottori 680-8550, Japan}
\address{(2) Core Research for Evolutional Science and Technology, 
Japan Science and Technology Agency (CREST-JST), Japan}
\address{(3) School of Computer Science, Tokyo University of Technology, 
Katakura-machi, Hachioji, Tokyo 192-0982, Japan}
\address{(4) Center for Research and Development of Higher Education, 
The University of Tokyo, 
Bunkyo-ku, Tokyo, 113-8656, Japan}
\address{(5) School of Information Science and Technology, 
Aichi Prefecture University, Nagakute-cho, Aichi 480-1198, Japan}
\address{(6) Department of Computational Science and Engineering, 
Nagoya University, Chikusa-ku, Nagoya 464-8603, Japan}

\ead{hoshi@damp.tottori-u.ac.jp}

\begin{abstract}
A linear-algebraic theory 
called \lq multiple Arnoldi method' is presented 
and realizes large-scale (order-$N$) electronic structure calculation
with generalized eigen-value equations.
A set of  linear equations, 
in the form of $(zS-H) \bm{x} = \bm{b}$,
are solved simultaneously with multiple Krylov subspaces.
The method is implemented 
in a simulation package ELSES (http://www.elses.jp) 
with tight-binding-form Hamiltonians.
A finite-temperature molecular dynamics simulation 
is carried out for metallic and insulating materials.
A calculation with $10^7$ atoms was realized by a workstation.
The parallel efficiency is shown upto 1,024 CPU cores.  
\end{abstract}

%\submitto{\JPCM}
\maketitle

\section{Introduction}

Large-scale electronic structure calculation, with $10^3$ atoms or more,
plays a crucial role in nano science and
is realized by an order-$N$ theory,
in which the computational cost is proportional to the system size.
References on the order-$N$ electronic structure theory 
can be found in a recent paper. \cite{TENG-2011} 
In this paper, a method, called \lq multiple Arnoldi method',  
is presented for generalized eigen-value equations,
or large-scale electronic structure theory with non-orthogonal (atomic) bases. 
The method is applicable both to metal and insulator and 
the molecular dynamics (MD) simulations
were carried out for upto ten-million-atom systems 
with tight-binding-form Hamiltonians.
The present method is a theoretical extension of 
a previous  one, 
the diagonalization method in the Krylov subspace,  
\cite{takayama_etal.04, HOSHI-2006-JPCM-NANOSTRC}, 
since the present method will be reduced to the previous one 
in the case with orthogonal bases.

This paper is organized as follows;
The theory is summarized in Sec.~\ref{SEC-MATH-FOUND}.
In Sec.~\ref{SEC-NUME}, 
numerical examples appear and 
the method is compared with 
several existing ones with  non-orthogonal bases. 
\cite{TENG-2011}
The summary is given in Sec.~\ref{SEC-SUMMARY}.

In this paper, the $j$-th unit vector is denoted as
$\bm{e}_j \equiv (0,0,0 \ldots , 1_j, 0,0,0, \ldots , 0_M)^{\rm T}$.
The inner product between the two vectors of 
$\bm{f} \equiv (f_1, f_2, \ldots)^{\rm T}, \bm{g} \equiv (g_1, g_2, \ldots)^{\rm T}$
is written as  
$\bm{f}^{\rm T} \bm{g} \equiv \sum_i f_i g_i $.
The unit matrix is denoted as $I$. 
The representation with atomic orbitals is considered and 
the suffix $i$ for a component of a vector indicates 
the composite suffix for atom and orbital ($s, p_x, p_y, p_z, ...$).

%%%%%%%%%%%%%%%%%%%%%%%%%%%%%%%%%%%%%%%%
\section{Theory \label{SEC-MATH-FOUND}}

A key concept of the present method is the Krylov subspace that is
defined as a linear (Hilbert) space of
\begin{eqnarray}
 K_\nu(A; \bm{b}) \equiv {\rm span} \{ \bm{b}, A\bm{b}, A^2\bm{b}, \ldots , A^{\nu-1}\bm{b} \}
\end{eqnarray}
with a given square matrix $A$ and a given vector $\bm{b}$.
Krylov subspace is a common mathematical foundation
for iterative  linear algebraic algorithms,
such as the conjugate-gradient (CG) algorithm.

A generalized eigen-value equation is written as
\begin{eqnarray}
H \bm{\phi}_k  = \varepsilon_k S \bm{\phi}_k.
\label{EQ-ORIG-EQ}
\end{eqnarray}
Here the Hamiltonian and overlap matrices are denoted as $H$ and $S$, respectively.
They are sparse real-symmetric $M \times M$ matrices
and $S$ is positive definite.  
The eigen levels and vectors are denoted as $\varepsilon_k$ and $\bm{\phi}_k$, respectively.

A basic equation for  large-scale electronic structure theory is the set of  linear equations
\begin{eqnarray}
 ( z S -H ) \bm{x}_j = \bm{e}_j
 \label{EQ-SHIFT-EQ}
\end{eqnarray}
among the unit vectors $\{ \bm{e}_{j=1},   \bm{e}_2,  \bm{e}_3, ....,  \bm{e}_M \}$.
A matrix element  of the Green's function, 
$G \equiv (zS-H)^{-1}$, \cite{NOTE-GREEN}
is given as $G_{ij}= \bm{e}_i^{\rm T}\bm{x}_j$.

In the present method, 
the solution of Eq.(\ref{EQ-SHIFT-EQ}) is given within 
the multiple Krylov subspace of
\begin{eqnarray}
{\cal L}_\nu^{(j)} = K_p(H; \bm{e}_j) \oplus K_q(H; S^{-1}\bm{e}_j),
\label{EQ-SUBSPACE-MULTI-KRY} 
\end{eqnarray}
where $p,q$ are positive integers and $\nu \equiv p+q$. 
The dimension of the subspace, $\nu$, is chosen to 
be much smaller than that of the original matrices. 
The case with $q=0$ is the generalized Arnoldi method in Ref.~\cite{TENG-2011}

The two initial vectors of 
$\bm{e}_j$ and $\bm{s}_j \equiv S^{-1}\bm{e}_j$ in Eq.~(\ref{EQ-SUBSPACE-MULTI-KRY})
satisfy a \lq duality' relation of $\bm{e}_j^{\rm T}S\bm{s}_j=1$.
A formulation with the dual vectors reduces 
An efficient numerical treatment of $S^{-1} \bm{e}_j$ is required
for a large-scale calculation, 
since the explicit matrix-inversion procedure of $S$ is costful, 
as the matrix-diagonalization procedure.
In the present method, 
the vector of $\bm{s}_j  : = S^{-1}\bm{e}_j$ is calculated
by an inner CG loop, in which 
the linear equation of $ S \bm{s}_j = \bm{e}_j$ is solved iteratively
with the standard CG method.
This inner loop converges fast, 
typically with $\nu_{\rm CG}=10-30$ iterations, 
since the overlap matrix is sparse and nearly equal to the unit matrix $(S \approx I)$.
\cite{TENG-2011}

The whole procedures are carried out in the following two stages.
First, the bases of the subspaces 
\begin{eqnarray}
{\cal L}_\nu^{(j)} \equiv {\rm span}\{ \bm{u}_1^{(j)},  \bm{u}_2^{(j)}, \ldots , \bm{u}_{\nu}^{(j)}\},
 \label{EQ-SUBS-MEMBERS}
\end{eqnarray}
are generated so as to satisfy the \lq $S$-orthogonality' 
($\bm{u}_m^{(j) {\rm T}} S \bm{u}_n^{(j)} = \delta_{mn}$);
With a given initial vector of $\bm{l}_{1} : = \bm{e}_{j} $, 
the $n$-th basis ($\bm{u}_n$), for $n \ge 1$ and $n \ne p+1$, 
is generated in the following three procedures;
\begin{eqnarray}
\bm{u}_{n} &  : =  &\frac{\bm{l}_{n}}{\sqrt{ \bm{l}_{n}^{t}S  \bm{l}_{n} }} 
 \label{EQ-REC-UN} \\
\bm{k}_{n} & : = & H \bm{u}_n 
 \label{EQ-KN-HUN} \\
\bm{l}_{n+1} &  : = & \bm{k}_{n}- \sum_{m=1}^{n} \bm{u}_{m} q_{mn}  
 \label{EQ-MGram-Schmidt} 
\end{eqnarray}
with $q_{mn} \equiv \bm{u}_{m}^{t} S  \bm{k}_{n}$.
The modified Gram-Schmidt procedure 
appear in Eq.~(\ref{EQ-MGram-Schmidt}), 
so as to satisfy the \lq S-orthogonality' of
$ \bm{u}_{m}^{t} S  \bm{k}_{n+1} = 0$ for $m=1,2, \ldots , n$.
For $n=p+1$, 
Eq.~(\ref{EQ-KN-HUN})  is replaced by 
\begin{eqnarray}
& & \bm{k}_{p+1} : =  S^{-1} \bm{e}_j.
\label{EQ-CALC-S-INV} 
\end{eqnarray}
The $S^{-1}$-vector multiplication in Eq.~(\ref{EQ-CALC-S-INV} )
is realized by the inner CG loop explained above.

Second, subspace eigen vectors $\bm{v}_{\alpha}^{(j)}$ 
($\subset {\cal L}_\nu^{(j)}$)
\begin{eqnarray}
\bm{v}^{(j)}_\alpha = \sum_n^{\nu} C_{n \alpha}^{(j)} \bm{u}_n^{(j)}
\label{EQ-SUB-EIG-VEC}
\end{eqnarray}
and subspace eigen levels $\veps^{(j)}_\alpha$ 
are introduced so that the residual vector
$\bm{r}^{(j)}_\alpha \equiv (H - \veps^{(j)}_\alpha S) \bm{v}^{(j)}_\alpha$
is orthogonal to  the subspace ($\bm{r}_{\alpha}^{(j)} \perp {\cal L}_\nu^{(j)}$). 
The above principle is known as Galerkin principle 
in numerical analysis. \cite{TEMPLATE}
Consequently, a standard eigen-value equation appears
with a reduced ($\nu \times \nu$)   Hamiltonian matrix of    
$ (H^{(j)})_{mn} \equiv \bm{u}_m^{(j) {\rm T}} H \bm{u}_n^{(j)} $.
The derived eigen-value equation is solved,
so as to determine $\veps^{(j)}_\alpha$ and $C_{n \alpha}^{(j)}$. 

The solution vector is determined as 
\begin{eqnarray}
\bm{x}_j(z) : =   G^{(j)}(z) \bm{e}_j 
\label{EQ-SOLUTION-SUB-GZ} 
\end{eqnarray}
where the matrix $G^{(j)}$, called \lq subspace Green's function', 
is defined as
\begin{eqnarray}
G^{(j)} \equiv \sum_\alpha 
\frac{ \bm{v}_\alpha^{(j)} \bm{v}_\alpha^{(j) {\rm T}} }{z - \veps_\alpha^{(j)}}.
\end{eqnarray}
The above calculation will be exact, 
when the subspace ${\cal L}_\nu^{(j)}$ comes to the complete space
($\nu \rightarrow M$).

The density matrix and the energy density matrix
\begin{eqnarray}
\rho_{ij}
 &: =&   \sum_{\alpha}^{\nu} f(\veps^{(j)}_\alpha) 
 \bm{e}_i^{\rm T}\bm{v}^{(j)}_\alpha \bm{v}^{(j) {\rm T}}_\alpha \bm{e}_j  \label{EQ-DM} \\
\pi_{ij}
 &: =&   \sum_{\alpha}^{\nu} f(\veps^{(j)}_\alpha) \veps^{(j)}_\alpha
 \bm{e}_i^{\rm T}\bm{v}^{(j)}_\alpha \bm{v}^{(j) {\rm T}}_\alpha \bm{e}_j
\end{eqnarray}
are calculated
where  the occupation number $f(\veps)$ is the Fermi-Dirac function
with the given values 
of the temperature (level-broadening) parameter and the chemical potential  $\mu$. 
The chemical potential 
is determined by the bisection method,
so that the total electron number is the correct one.

The electronic structure energy ($E_{\rm elec}$) and
its derivative with respect to the $K$-th atom position ($\bm{F}_K $) 
are required for a MD simulation.
They are decomposed into the partial sums as
\begin{eqnarray}
E_{\rm elec} &\equiv&   {\rm Tr}[\rho H] = \sum_{j}E_{\rm elec}^{(j)} \label{EQ-ENE} \\
\bm{F}_K &\equiv& - \frac{\partial E_{\rm elec}}{\partial \bm{R}_K} 
 =  \sum_{j} \bm{F}_{K}^{(j)} \label{EQ-FORCE},
\end{eqnarray}
where the partial sums are defined by 
\begin{eqnarray}
E_{\rm elec}^{(j)} &\equiv& \sum_{i} \rho_{ij} H_{ji} \\
 \bm{F}_{K}^{(j)} & \equiv & - \sum_{i} \left\{ \rho_{ij} \frac{\partial H_{ji}}{\partial \bm{R}_K} 
 +  \pi_{ij} \frac{\partial S_{ji}}{\partial \bm{R}_K} \right\}. 
 \label{EQ-FORCE-PART}
\end{eqnarray}
The  components of $\rho_{ij}$ or $\pi_{ij}$ are required
only for the selected $(i,j)$ pairs 
that satisfy $H_{ij} \ne 0$ or $S_{ij} \ne 0$, respectively.
The value of $\bm{F}_{K}^{(j)}$ in Eq.~(\ref{EQ-FORCE-PART}) 
is contributed only within a local region
where the atom positions of the $i$-th and $j$-th bases 
are equal to or near the $K$-th atom position ($\bm{R}_K$),
because the value of $(\partial H_{ji}/ \partial \bm{R}_K)$
or $(\partial S_{ji}/ \partial \bm{R}_K)$ is non-zero
only for the local region.

The calculation work flow is summarized as
\begin{eqnarray}
& & \{ \{ \bm{u}_n^{(j)} \} \Rightarrow 
\{ \bm{v}^{(j)}_\alpha , \veps^{(j)}_\alpha \} \}_j 
\Rightarrow  ({\rm bisection}) \Rightarrow  \mu \nonumber \\
&\Rightarrow &   \{ f(\veps^{(j)}_\alpha) \}_j 
\Rightarrow  \{ \{ \rho_{ij}, \pi_{ij} \} \Rightarrow  \{ E_{\rm elec}^{(j)}, \{  F_{K}^{(j)}\} \}\}_j 
 \label{EQ-WORK-FLOW}
\end{eqnarray}
where  the procedures in a curly parenthesis $\{ \cdot  \cdot  \cdot  \cdot \}_j$ are carried out
independently among the running index $j$, as a parallel computation. 
In the bisection procedure, 
the total electron number with a trial value of the chemical potential
is summed up among the bases 
and the summation is parallelized with the basis index $j$. 

Several calculations with 
the charge-self-consistent (CSC) formulation
\cite{ELSTNER-1998-CSC} were also carried out.
At each MD step, 
an iterative loop is required for the self consistency of the change distribution.
Since the overlap matrix is unchanged within the iterative loop,
the inner CG loop for $ S^{-1} \bm{e}_j$ is required 
only once  at one MD step
and gives a tiny fraction of the total computational cost.

%%%%%%%%%%%%%%%%%%%%%%%%%%%%%%%%%%%%%%%%%%
\begin{figure}[htbp] 
\begin{center}
  \includegraphics[width=8cm]{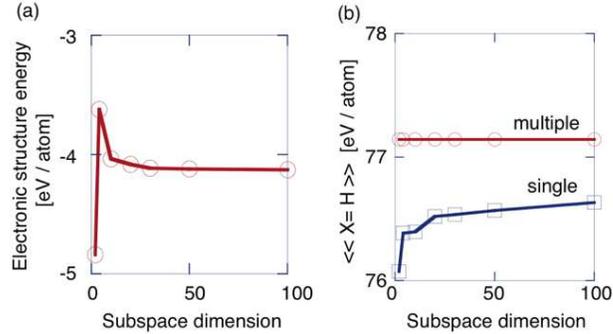}
  % \vspace{8cm}
\end{center}
\vspace{-5mm}
\caption{\label{fig-graph-energy-sum} 
Numerical example of solid gold with 864 atoms. 
(a) The electronic structure energy in the multiple subspace method, 
calculated with the subspace dimension of $\nu = 2, 4, 10, 20, 30, 50, 100$.
(b) The estimated energy value from the whole spectrum
 in the multiple subspace method (red line) and
in the single subspace method (blue line), 
calculated with the subspace dimension of $\nu = 2, 4, 10, 20, 30, 50, 100$.
}
\end{figure}
%%%%%%%%%%%%%%%%%%%%%%%%%%%%%%%%%%%%%%%%%%

%%%%%%%%%%%%%%%%%%%%%%%%%%%%%%%%%%%%%%%%%%
\begin{figure}[htbp] 
\begin{center}
  \includegraphics[width=8cm]{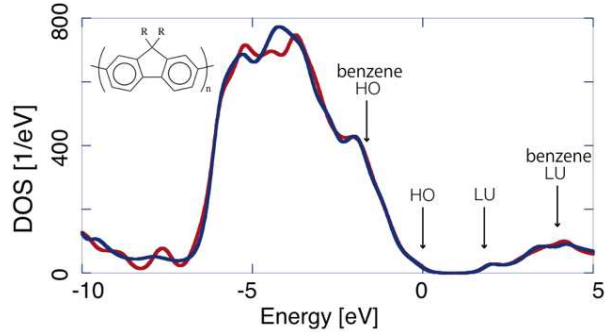}
  % \vspace{8cm}
\end{center}
\vspace{-5mm}
\caption{\label{fig-DOS-PFO} 
The density of states of an amorphous-like conjugated polymer
(poly-(9,9 dioctyl-fluorene)), calculated
by the present method (blue line) and 
by the exact diagonalization method (red line). 
The HO and LU  levels are indicated by arrows. 
The HO and LU levels of benzene are also indicated by arrows.
The inset shows the monomer unit with $R \equiv$ C$_8$H$_{17}$.
}
\end{figure}
%%%%%%%%%%%%%%%%%%%%%%%%%%%%%%%%%%%%%%%%%%

%%%%%%%%%%%%%%%%%%%%%%%%%%%%%%%%%%%%%%%%
\section{Examples and discussion  \label{SEC-NUME}}

Several numerical examples are calculated by the multiple Arnoldi method. 
We choose $p=q=\nu/2$ ($\nu$ : even) in the following calculations,
except where indicated,
so as to investigate the examples,
systematically among different values of the subspace dimension ($\nu$),
with a significant contribution by the second term in Eq.~(\ref{EQ-SUBSPACE-MULTI-KRY}).

Figure \ref{fig-graph-energy-sum}(a) shows  
the electronic structure energy $E_{\rm elec}$ 
for bulk gold with 864 atom.
The tight-binding-form Hamiltonian in Ref. \cite{NRL} was used
and contains $s$, $p$ and $d$ orbitals. 
The electronic structure energy was calculated
with the subspace dimensions of $\nu = 2, 4, 10, 20, 30, 50, 100$.
The calculated energy agrees for $\nu =30, 50, 100$
within deviations less than 0.01 eV per atom. 
In general, 
the use of the multiple Krylov subspaces ($p,q \ge 1$)
reproduces several properties.
(i) In the fully filled limit ($f(\varepsilon) \rightarrow 1$), 
a physical quantity is contributed by all the eigen states 
and is expressed by
\begin{eqnarray}
  \langle  \langle X \rangle \rangle
  \equiv  \sum_k \bm{\phi}_k^{\rm T} X \bm{\phi}_k 
  = {\rm Tr} [S^{-1} X].
  \label{EQ-PHYS-FULL-OCC}
\end{eqnarray}
with a real-symmetric matrix $X$. 
One can prove the fact that Eq.~(\ref{EQ-PHYS-FULL-OCC}) holds exactly, 
if $q \ge 1$ (or $S^{-1} \bm{e}_j \subset {\cal L}^{(j)}$).
\cite{NOTE-PROOF}
Figure ~\ref{fig-graph-energy-sum}(b)
confirms the fact numerically in the case of $X=H$.
(ii) One can also prove that
the equivalence of the two expressions of the band structure energy
(${\rm Tr}[\rho H] = {\rm Tr}[\pi S]$) \cite{TENG-2011} holds exactly, if $p \ge 1$
 (or $\bm{e}_j \subset {\cal L}^{(j)}$). 
The equivalence  was confirmed numerically (not shown).

A MD simulation for a semiconducting system  
was carried out with $\nu=30$ 
for an amorphous-like structure of a conjugated polymer, 
poly-(9,9 dioctyl-fluorene) with 2076 atoms. 
\cite{PFO}  
The simulation was carried out with 
the tight-binding Hamiltonian of a modified extended H\"{u}ckel type 
in Ref.~\cite{ASED-CALZAFFERI}. 
The results for the monomer and dimer 
agree reasonably to those 
by the {\it ab initio} calculation of Gaussian$^{\rm (TM)}$
with the B3LYP functional and the 6-311G(d,p) basis set.
Detailed data by the present method are added here
with those by the {\it ab initio} calculation in the parentheses; 
The valence band width $W$ and the band gap $\Delta$ are
$W=18.5$ eV (18.3 eV) and $\Delta =4.25$ eV (4.91 eV) in the monomer
and $W=19.0$ eV (18.8 eV)  and $\Delta = 3.58$ eV (4.10 eV) in the dimer.
The two monomers in the dimer are twisted along the main chain
and the twisting angle $\theta$ is 
$\theta=37.3 ^\circ$ (40.6 $ ^\circ$).
As a technical detail in large-scale calculations, 
the real-space projection method was used
and is explained in Appendix of Ref. ~\cite{HOSHI-2006-JPCM-NANOSTRC} 
In short, the Krylov subspace is generated by a Hamiltonian 
projected in real space, $H^{(j)} \equiv P^{(j)} H P^{(j)}$, instead of the original one $H$, 
where the projection operator $P^{(j)}$ projects a function onto 
the spherical region whose center is located at the
atomic position of the $j$ th atomic basis.
The projection radius is determined for each basis $\bm{e}_j$,
so that the region contains $N_{\rm RP}$ atoms or more.
The same technique is used also for the overlap matrix. 
The value of $N_{\rm RP}$ is an input parameter 
and is set to $N_{\rm RP}=100$.

Figure \ref{fig-DOS-PFO} shows
the density of state (DOS), calculated from the Green's function, 
for the amorphous-like conjugated polymer.
The calculation of DOS requires a finer calculation conditions
($\nu$ = 300 and $N_{\rm RP} = 1000$) than that for the density matrix,
since the DOS profile is an energy resolved quantity.
The result by the exact diagonalization method is also shown and
one finds that the present method reproduces the overall spectrum precisely.
Moreover, 
when the eigen levels are assumed to be non-degenerated,
the calculated Green's function can be decomposed into
the contributions of individual eigen states 
and the individual eigen levels can be estimated. 
 \cite{NOTE-DOS-DECOMP} 
For example, 
the highest-occupied (HO) and lowest occupied (LU) levels were estimated
and are indicated by the arrows in Fig.~\ref{fig-DOS-PFO}.
These values agree excellently, within less than 3 meV, with those in the exact diagonalization method.
The agreement is also found on a couple of levels near the HO and LU levels.
It is noteworthy that 
a state located near a band edge, such as HO and LU states,
is reproduced with a smaller subspace dimension ($\nu$) than 
one located within the band, 
as a  general property of the subspace theory. \cite{TAKAYAMA2006}

A MD simulation was carried out  also 
for a gold nanowire, a metal. 
The same conditions of $\nu$ and $N_{\rm PR}$ were used as in the polymer simulation.
The simulation by the present method reproduces
the formation process of helical gold nanowire,
as ones by the exact diagonalization method. 
\cite{IGUCHI-2007-PRL-HELICAL, HOSHI-2009-JPCM-HELICAL}

%%%%%%%%%%%%%%%%%%%%%%%%%%%%%%%%%%%%%%%%%%
\begin{figure}[htbp] 
\begin{center}
  \includegraphics[width=8cm]{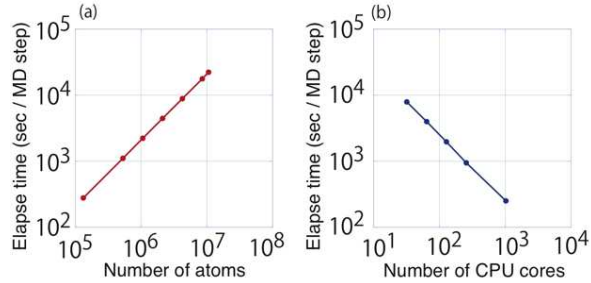}
  % \vspace{8cm}
\end{center}
\vspace{-5mm}
\caption{\label{fig-order-n} 
Calculation of  the amorphous-like polymer.
(a) Order-$N$ scaling property upto 10$^7$ atoms.
(b) Parallel efficiency with 10$^7$ atoms upto 1,024 CPU cores. 
}
\end{figure}
%%%%%%%%%%%%%%%%%%%%%%%%%%%%%%%%%%%%%%%%%%

A high computational efficiency is shown 
for the present method among calculations of  the conjugated polymer.
When the system with 2076 atoms was calculated 
by a work station with two six-core Xeon CPUs (X5650),  
the present method consumes 2.6 seconds per MD step and is
faster, approximately by ten times, than the exact diagonalization method.
A higher efficiency is obtained for a larger system,
since the present method consumes an $O(N)$ cost for an $N$-atom system,
whereas the exact diagonalization method consumes an $O(N^3)$ cost.  
Figure \ref{fig-order-n} (a) shows that 
the calculation has the order-$N$ scaling property with upto 10,629,120 atoms.  \cite{NOTE-WS}
Figure \ref{fig-order-n} (b) shows the parallel efficiency of the present method 
with the ten-million-atom system, among 32 - 1,024 cores. 
The MPI/OpenMP hybrid parallelism was carried out
by quad-core Xeon CPUs (X5570) of SGI Altix ICE 8400EX. 
The calculation did not work with smaller numbers of cores,
because of the insufficient memory.  
The parallel efficiency is almost ideal,
since the measure for the efficiency is obtained as 
$\alpha \equiv T(32)/T(1024) \times (1024/32)=0.994$,
where $T(n)$ is the elapse time with $n$ cores per MD step.
The dominant part of the elapse time is that for 
the electronic structure calculation with the work flow of Eq.~(\ref{EQ-WORK-FLOW}) 
and the rest parts contain the file IO and other procedures.
The parallel efficiency only for the electronic structure calculation
is higher ($\alpha = 1.00$) than that for the whole elapse time.
The high parallel efficiency  appears,
because only vector quantities in small data sizes, 
such as the force on atoms (${\bm{F}_I}$), 
are communicated among the nodes.
Matrix quantities ($H, S, \rho, \pi$) in much larger data sizes 
are not communicated among the nodes; \cite{GESHI}
The required elements of $H$ and $S$ are calculated redundantly among the nodes
and the elements of $\rho_{ij}$ and $\pi_{ij}$ are  calculated and used 
only within the procedures  parallellized by the index $j$, 
as shown in Eq.~(\ref{EQ-WORK-FLOW}).

Finally, 
the efficiency of the present method is compared with 
the other subspace methods proposed 
in Ref.~\cite{TENG-2011} or the references therein;
generalized shifted conjugate-orthogonal conjugate gradient (gSCOCG) method 
and
generalized Lanczos (gLanczos) method.
In these methods, 
a Krylov subspace of $K_\nu(S^{-1}H; \bm{b}) $
is used for an initial vector $\bm{b}$.
Then the inner CG loop for the $S^{-1}$-vector multiplication
appears at every step of the recurrence relation,
unlike Eq.~(\ref{EQ-KN-HUN}), and 
requires $\nu_{\rm CG}$ time matrix-vector multiplications. 
The present method gives, therefore, a faster performance,
when  the computational cost is dominated 
by the matrix-vector multiplications,
as those in the MD simulations of the present paper.
For example, the measured computational time 
in the gLanczos method with the same subspace dimension ($\nu$)
is six times larger than 
that of the present one 
or the benchmark data with $1.4 \times 10^5$ atoms in Fig.~\ref{fig-order-n} (a).
The faster performance of the present  method, however,
may not hold, 
when the number of the subspace dimension ($\nu$) 
is much larger than those in the present paper ($\nu \rightarrow M$)
and the cost is dominated by the procedure of 
calculating the subspace eigen vectors of Eq.~(\ref{EQ-SUB-EIG-VEC}) for the given reduced matrix. 
This is because, in the present  method,
the reduced $\nu \times \nu$ matrix is dense  and
the procedure consumes an $O(\nu^3)$ cost.
The subspace methods with the subspace of $K_\nu(S^{-1}H; \bm{b})$ 
avoid  the $O(\nu^3)$ cost, since the reduced matrix is tridiagonal. 
In conclusion,
one should use the present method first 
with a moderate number of the subspace dimension ($\nu =10^1$-$10^{2}$)
and, if one finds a serious demand for a much larger number of $\nu$,
one may use another method explained above.  
In addition, the gSCOCG and gLanczos methods have several advantages; 
The energy momenta are conserved by the $\nu$-th order
in the two methods and
the calculation by the gSCOCG method is robust against numerical rounding errors, 
even without the explicit modified Gram-Schmidt orthogonalization procedure or  
the long recurrence of Eq.~(\ref{EQ-MGram-Schmidt}). \cite{TENG-2011}
The absence of the long recurrence saves both the CPU time and memory costs,
among the calculation with a large subspace dimension.

%%%%%%%%%%%%%%%%%%%%%%%%%%%%%%%%%%%%%%%%

\section{Summary \label{SEC-SUMMARY}}

The \lq multiple Arnoldi method' is presented 
for large-scale (order-$N$) electronic structure calculation
with non-orthogonal bases. 
The test calculations were carried out with upto $10^7$ atoms.
The present paper shows the potential of the present method,
since the method is applicable both to metals and insulators 
and shows an ideal parallel efficiency.
The method is implemented in a simulation package ELSES (http://www.elses.jp).

\section*{Acknowledgement}

This research was supported partially by Grant-in-Aid 
(KAKENHI, No. 20103001-20103005, 23104509, 23540370), 
from the Ministry of Education, Culture, Sports, Science and Technology (MEXT) of Japan.
The parallel computation in Fig.~\ref{fig-order-n} (b)
was carried out using the supercomputer
of the Institute for Solid State Physics, University of Tokyo.
The supercomputers at the Research Center for 
Computational Science, Okazaki were also used. 
The authors thank Y. Zempo (Hosei University) and M. Ishida (Sumitomo Chemical Co., Ltd) 
for providing the structure model of the amorphous-like polymer.

%%%%%%%%%%%%%%%%%%%%%%%%%%%%%%%%%%%%%%%%%%%%

\appendix

%%%%%%%%%%%%%%%%%%%%%%%%%%%%%%%%%%%%%%%%%%%%%%%%


\section*{References}
\begin{thebibliography}{100}

\bibitem{TENG-2011}
Teng H, Fujiwara T, Hoshi T, Sogabe T, Zhang S-L, and Yamamoto S 2011
Phys. Rev. B {\bf 83} 165103

\bibitem{takayama_etal.04} 
Takayama R, Hoshi T, and Fujiwara T 2004
J. Phys. Soc. Jpn. {\bf 73} 1519

\bibitem{HOSHI-2006-JPCM-NANOSTRC} 
Hoshi T and Fujiwara T 2006
J. Phys.: Condens. Matter {\bf 18} 10787

\bibitem{NOTE-GREEN}
One should distinguish the present definition of the Green's function, 
from that of $G=S(zS-H)^{-1}S$ 
in Ref.~\cite{TENG-2011}
and other papers.

\bibitem{ARTACHO-1991}
Artacho E and Mil\'{a}ns del Bosch L 1991
Phys. Rev. A {\bf 43} 5770 

\bibitem{TEMPLATE}
Bai Z, Demmel J, Dongrarra J,
Ruhe A, and van der Vorst H 2000 
{\it  Templates for the Solution of Algebraic Eigenvalue Problems},
SIAM, Philadelphia

\bibitem{ELSTNER-1998-CSC}
Elstner M, Porezag D, Jungnickel G, Elsner J, 
Haugk M, Frauenheim Th, Suhai S and Seifert G 1998 
Phys. Rev. B {\bf 58} 7260 

\bibitem{NRL}
Mehl M J and Papaconstantopoulos D A 1996
Phys. Rev. B, {\bf 54} 4519; 
Kirchhoff F,  Mehl M J, Papanicolaou N I,
Papaconstantopoulos D A and Khan F S 2001
Phys. Rev. B {\bf 63}, 195101;
Papaconstantopoulos D A and Mehl M J 2003
J. Phys.: Condens. Matter {\bf 15} R413 

\bibitem{NOTE-PROOF}
The proof is based on a \lq projection' theorem:
In the  fully filled limit,
the density matrix of Eq.~(\ref{EQ-DM})  is reduced to 
$\Omega^{(j)} \equiv \sum_n \bm{u}_n^{(j)}  \bm{u}_n^{(j) {\rm T}}$.
If a vector $\bm{\gamma}^{(j)}$ is included in the subspace 
($\bm{\gamma}^{(j)} \subset {\cal L}^{(j)}$), 
the \lq projection' theorem of
$\Omega ^{(j)} S \bm{\gamma}^{(j)} = \bm{\gamma}^{(j)}$
holds. 


\bibitem{PFO}   
See experimental papers, such as
Chen S H, Chou H L,  Su A C, and  Chen S A 2004 
Macromolecules {\bf 37} 6833

\bibitem{ASED-CALZAFFERI}
Calzaferri G and Rytz R 1996 
J. Phys. Chem. {\bf 100} 11122

\bibitem{NOTE-DOS-DECOMP}
Each eigen level is assigned from 
the inverse function of the integrated density of states,$\eta = \eta(n)$, as follows;
The integrated DOS is assumed to be the integration of 
smoothed delta functions of non-degenerated levels ($\sum_k \delta(\varepsilon - \varepsilon_k)$).
The energy integration in the region of 
$\eta(k-1) < \varepsilon < \eta(k)$ 
is  assigned to be the contribution of the $k$-th eigen state.
The $k$-th eigen level is estimated to be
$\varepsilon_k : = \eta(k-1/2)$
as the central peak position of the smoothed delta function.


\bibitem{TAKAYAMA2006}
Takayama R,  Hoshi T, Sogabe T, Zhang S-L and Fujiwara T 2006  
Phys. Rev. B {\bf 73} 165108

\bibitem{IGUCHI-2007-PRL-HELICAL} 
Iguchi Y, Hoshi T and Fujiwara T 2007 
Phys. Rev. Lett. {\bf 99} 125507

\bibitem{HOSHI-2009-JPCM-HELICAL}
 Hoshi T and Fujiwara T 2009 
 J. Phys.: Condens. Matter {\bf 21} 272201

\bibitem{NOTE-WS}
A ten-million-atom calculation was realized for a bulk silicon
by a perturbation method of the Wannier state
(Fig.10 of T.~Hoshi, Y.~Iguchi and T.~Fujiwara, Phys. Rev. B {\bf 72}, 075323 (2005)).
Its applicability, however, is severely limited, unlike the present method, 
since the Wannier states are constructed from the occupied states 
and the method is applicable only to insulating systems.
Moreover the perturbation theory
requires reliable initial states as unperturbed wavefunctions.

\bibitem{GESHI}
Geshi M,  Hoshi T and Fujiwara T 2003 
J. Phys. Soc. Jpn. {\bf 72} 2880



\end{thebibliography}
\end{document}